\documentclass{elsart}
\usepackage{graphicx,natbib,amssymb,lineno}

\makeatother

\begin{document}
\begin{frontmatter}
\title{ Magellanic Stream: A Possible Tool for Studying Dark Halo Model}

\author{Hosein Haghi, Akram Hasani Zonooz}
\address{Department of Physics, Institute for Advanced Studies in Basic Sciences (IASBS), P. O. Box 45195-1159, Zanjan, Iran }
\author{Sohrab Rahvar\corauthref{cor}}
\address{Department of Physics, Sharif University of
Technology, P.O. Box 11365-9161, Tehran, Iran}
\corauth[cor]{Corresponding author.}
\ead{rahvar@sharif.edu}

\begin{abstract}
We model the dynamics of Magellanic Stream with the ram-pressure
scenario in the logarithmic and power-law galactic halo models and
construct numerically the past orbital history of Magellanic clouds
and Magellanic Stream. The parameters of models include the
asymptotic rotation velocity of spiral arms, halo flattening, core
radius and rising or falling parameter of rotation curve. We obtain
the best fit parameters of galactic models through the maximum
likelihood analysis, comparing the high resolution radial velocity
data of HI in Magellanic Stream with that of theoretical models. The
initial condition of the Magellanic Clouds is taken from the six
different values reported in the literature. We find that oblate and
nearly spherical shape halos provide a better fit to the observation
than the prolate halos. This conclusion is almost independent of
choosing the initial conditions and is valid for both logarithmic
and power-law models.
\end{abstract}
\begin{keyword}
galaxies: Magellanic Clouds \sep Galaxy: structure \PACS 98.56.Tj

\end{keyword}

\end{frontmatter}
\section{Introduction}
Analysis of rotation curve of spiral galaxies reveals a faster
rotation curve at the outer parts of spiral arms compare to what is
expected based on the distribution of visible matter. Hence invoking
a halo shape dark component for galaxies is essential to explain the
dynamics of galaxies (Bosma 1981; Rubin and Burstein 1985). Despite
using a large number of techniques, the shape of the galactic halo
is not well known and is still in debate. It is even not clear if
the halo is prolate or oblate. Most halo models expect significant
deviations from the spherical symmetry (Frenk et al. 1988; Dubinski
and Carlberg 1991; Olling and Merrifield 2000 and Ruzicka et al.
2007). An accurate and reliable determination of the shape and mass
distribution of the galactic halo requires the analysis of the
kinematic properties of several halo objects (i.e. satellite
galaxies and globular clusters). Another promising method is
studying the stellar and gaseous streams which produced by
disruption and accretion of small galaxies orbiting around the Milky
Way (MW). The shape and the kinematics of such streams are strongly
influenced by the overall properties of the underlying potential.
There are several examples of such streams which have been
discovered in recent years (Odenkirchen et al. 2001; Grillmair \&
Dionatos 2006; Ibata et al. 2001; Newberg et al. 2002; Majewski et
al. 2003). The most earliest discovered stream is the Magellanic
Stream (MS) which is a gaseous streams associated to the Magellanic
Clouds (MCs).

The MS is a narrow band of neutral hydrogen clouds lies along a
great circle from $(l=91^{\circ},b=-40^{\circ})$ to
($l=299^{\circ},b=-70^{\circ}$), started from the MCs and oriented
towards the south galactic pole. The radial velocity and the column
density of MS has been measured by several groups
\cite{mat87,bru05}. Observations show that the radial velocity of MS
with respect to the Galactic center changes from $0~kms^{-1}$ at the
nearest distance from the center of MCs to $-200~ kms^{-1}$ at the
tail of stream. Another feature of this structure is that the column
density falls about two orders of magnitude along the stream. The
two main competing scenario for explaining the origin of MS are
based on the tidal interaction and the ram pressure stripping which
predict different ages and spatial distributions for MS. In the
tidal model the idea is the gravitational stripping of gas from the
MCs due to the strong tidal interaction of LMC and SMC with Galaxy
at their perigalactic passage (Murai and Fujimoto 1980; Lin \&
Lynden-Bell 1982; Gardiner \& Noguchi 1996; Connors et al. 2006).
This leads to the formation of tidal tails emanating from the
opposite side of MCs. In the tidal scenario, although the observed
radial velocity profile of the stream has been modeled remarkably
well, the smooth HI column density distribution does not match with
the observations and the expected stars in the stream have not been
observed yet. The other hypothesis is the ram pressure stripping of
gas from the MCs by an extended halo (Moore \& Davis 1994; Heller \&
Rohlfs 1994; Sofue 1994). In this model, there is a diffused halo
around the Galaxy which produces a drag on the gas within the MCs
and causes material to escape and form a trailing gaseous stream.
Compared to the tidal scenario, the ram pressure model allows a
better production of HI, compatible with the column density of MS.

In this work, we study the dynamics of MS for two generic galactic
models to constraint their parameters based on detailed HI
observations of MS \cite{bru05}. The main problem with modeling the
interaction of the MCs with the MW is the uncertainty in the initial
condition of MCs.
%and parameters of galactic models. Since a
%complete investigation of the full parameter space is a time
%consumer task, some assumptions are essential.
Here we do calculations for six different initial conditions of MCs
which have been reported in the literatures (Table \ref{tmodel}).

The paper is organized as follows: In section \ref{S2} we introduce
the potential models of MW and MCs. In section \ref{S3} we study the
dynamics of MS in various Galactic models and compare them with the
observations. The best parameters of model is obtained through the
maximum likelihood analysis in section \ref{S4}. The conclusion is
given in section \ref{S5}.

%========================================================================================================

\section{Models for Milky Way and Magellanic Clouds}
\label{S2}
\subsection{Milky Way}
The galactic model for the gravitational potential of an spiral
galaxy consists of a disk, bulge and halo components. For the halo
component of MW we take two sets of models so-called "power-law" and
"logarithmic" potentials. The axisymmetric power-law and logarithmic
models for the galactic halo are given by the following potentials
\cite{eva93}:
\begin{equation}
\label{phiL}
 \Phi_{L}=  -{\frac{1}{2}}{{V_a}^2\log{({R_c}^2 +
R^2 +z^2q^{-2})}},
\end{equation}
and
\begin{equation}
\label{phiP}
 \Phi_P= \frac{{\Psi_a}{R_c}^{\beta}}{({R_c}^2 +
R^2 +z^2q^{-2})^{\beta/2}},
\end{equation}
%where $\Psi_a =\frac{v_a^2}{\beta}$

where $\Psi_a =\frac{V_a^2}{\beta}$ is the central potential, $R_c$
is the core radius and $q$ is the flattening parameter. $q=1$
represents a spherical halo and $q \neq1$ gives an elliptical shape.
The parameter $\beta$ determines whether the rotation curve
asymptotically rises $(\beta <0)$, falls $(\beta>0)$ or is flat
$(\beta = 0)$. These models can reproduce the observed rotation
curve of MW.

The galactic disc and bulge are the luminous components of the MW.
For simplicity in calculation at the distances $r>50 kpc$, we take
the gravitational contribution of bulge and disk are as point like
objects with the masses of $3.5\times10^{10} M_{\odot}$ and
$1.2\times10^{11} M_{\odot}$, respectively \cite{hel04,read05}.

\subsection{Magellanic Clouds}
For LMC and SMC we adopt the Plummer potential as:
\begin{equation}
\Phi_{L,S}(r)= \frac{Gm_{L,S}}{[( {\mathbf{r}} -
{\mathbf{r}}_{L,S})^{2}+ K_{L,S}^{2}]},
\end{equation}
where ${\mathbf{r}_{L,S}}$ are the position of the Clouds relative
to the MW center, and $K_{L}=3 kpc$ and $K_{S}=2 kpc$ are the core
radii. We assume a mass of $m_{L}= 2\times 10^{10}M_{\odot}$ for the
LMC, and $m_{S}= 3\times 10^{9}M_{\odot}$ for the SMC.
%========================================================================================================
\section{Dynamics of the Magellanic system }\label{S3}

The simulation of MCs requirers the knowledge of 3D orbits of LMC
and SMC. We observe only a 2D projection on the sky at the present
time so 3D distribution of positions and velocities are partly
unknown. While the spatial location of MCs is well known, the the
velocity field corresponds to this structure due to the uncertainty
in the transverse velocity is less constrained. It is therefore very
difficult to construct a three-dimensional orbit that matches all
observations.

In this section we use the RK4 method\footnote{Runge-Kutta} to
calculate the equations of motion of MCs. We integrate the equation
of motion numerically and extract the dynamics of MCs by
backtracking orbits from their current positions and velocities to
$t=-5 Gyr$. The equations of motion of the Clouds is given by:
\begin{equation}
\label{lmc}
\frac{d^{2}\mathbf{r}_{L,S}}{dt^{2}}=\frac{\partial}{\partial
r_{L,S}}[\phi_{S,L}(|\mathbf{r}_{L,S}-\mathbf{r}_{S,L}|)+
\phi_{G}(|\mathbf{r}_{L,S}|)] + \mathbf{f}_{L,S} +
\textbf{F}_{drag},
\end{equation}
where $\phi_{S,L}$ represents mutual gravitational interaction of
LMC and SMC, $\phi_G$ is the gravitational potential of Galaxy,
$\textbf{f}$ is dynamical friction force, $\mathbf{r}$ is the
distance of each clouds from the Galactic center and
$\textbf{F}_{drag}$ is the hydrodynamical drag force. At the first
step, before studying the dynamics of MS we examine the effect of
various friction forces on the dynamics of MCs.

\subsection{Evaluation of friction forces}
In this part, we evaluate the contribution of friction forces in the
dynamics of MCs. We consider a simple isothermal halo where the
potential is given by $\phi_G(r)= -{{v_c}^2\ln{r}}$,
%\begin{equation}
%\phi_G(r)= -{{V_c}^2\ln{r}},
%\end{equation}
and $v_{c}=220 kms^{-1}$ \cite{con05}. In this calculation we take
the initial conditions of MCs as reported in HR94 \cite{hel94}, see
Table \ref{tmodel}. In this calculation we use the Cartesian
coordinate system where the x-axis is defined in the direction from
the sun to Galactic center, the y-axis is points towards the
direction of the circular motion of Sun around the Galaxy, and the
z-axis is directed to the northern Galactic pole.
%\begin{equation}
%\label{init} \mathbf{v}_L=(-10.06; -287.09; 229.73),\hspace{0.5cm}
%\mathbf{r}_{L}=(-0.85; -40.85; -27.95)
%\end{equation}
%\begin{equation}
%\label{init1} \mathbf{v}_{S}=(-32.00; -295.16; 196.46),
%\hspace{0.5cm} \mathbf{r}_{S}=(15.0; -36.0; -41.0).
%\end{equation}

\begin{figure}
\begin{center}
\includegraphics[width=140mm,height=110mm]{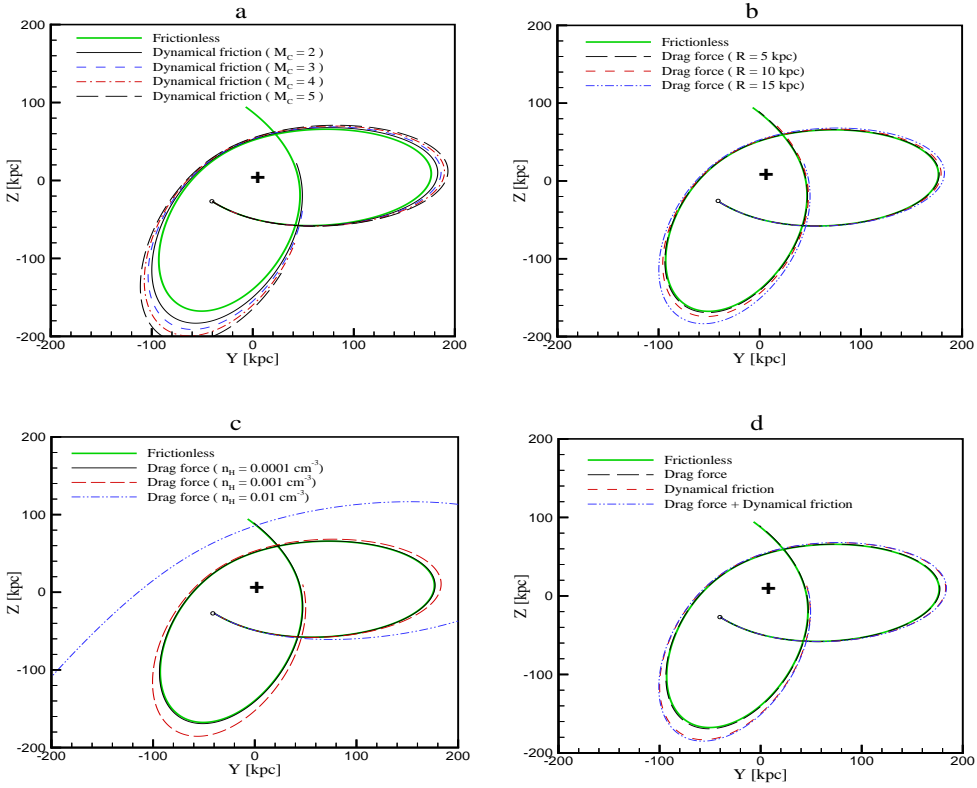}
\end{center}
\caption{ The effect of friction forces on the orbit of MCs. (a) The
effect of dynamical friction for different value of MCs's mass. (b)
The effect of hydrodynamical drag force on the orbit of MCs for
different value of MCs's size. (c) The effect of hydrodynamical drag
force on the orbit of MCs for different value of halo density. (d)
Comparison of orbits for different types of friction forces with
that of frictionless orbit. Here is shown that the effect of
dynamical friction is more significant than the hydrodynamical drag
force. The duration of evolution is $5 Gyr$. The mass of MCs is in
terms of $10^{10}M_{\odot}$. The initial condition selected as HR94.
The plus sign denotes the Galactic position at (0,0,0). The black
spot denotes the present position of MCs. } \label{d1}
\end{figure}

(a) {\it Dynamical friction}: Due to the Galactic extended massive
halo, the satellite galaxies passing through halo will be slowed
down by the gravitational force of the halo so-called dynamical
friction. This force makes MCs to spiral slowly toward the Galactic
center. We adopt the following standard form of dynamical friction
for the halo \cite{bin87}:
\begin{equation}
\mathbf{f}_{L,S}= 0.428 \ln\Lambda \frac{Gm_{L,S}^2}{r^{2}}
\frac{\mathbf{v}_{L,S}}{v_{L,S}},
\end{equation}
where $ \ln\Lambda\sim 3$ is the Coulomb logarithm. To evaluate the
strength of the dynamical friction force versus the gravitational
force of the halo, we compare their corresponding time scales. The
dynamical time scale can be given by the orbital period as $T_{dyn}
\sim r/v_c $, where $v_c$ is the circular velocity and $r$ is the
distance of MCs from the MW. On the other hand the time scale of
dynamical friction is in order of $T_{fric}\sim r^2 v_c/{GM_{MCs}}$,
where $M_{MCs}$ is the mass of the Magellanic Clouds. The ratio of
$T_{dyn}/T_{fric}\sim M_{MCs}/M_{MW}$ shows that $T_{fric}\gg
T_{dyn}$. Figure~(\ref{d1}a) shows the effect of dynamical friction
on the orbit of MCs where the spiraling of clouds get faster by
increasing the mass of clouds.

\begin{table*}
\begin{center}
\begin{tabular}{|c|c|c|c|c|c|}
\hline
work         &3D $v(x,y,z)$               & $|v|$& $v_{tan}$& $v_{rad}$& $r(x,y,z)$      \\
\hline$GN96$   &(-5,-226,194)                   & 297        & 287            & 82             & (-1.0,-40.8,-26.8)   \\
      $HR94$   &(-10.06,-287.09,229.73)         & 367.83     & 351.81         & 107.37         & (-0.85,-40.85,-27.95)\\
      $K2$     &(-91,-250,220)                  & 345        & 333            & 92             & (-0.8,-41.5,-26.9)   \\
      $K1 Mean$&$(-86^{+12}_{-12},-268^{+11}_{-11},252^{+16}_{-16})$ & $378^{+18}_{-18}$ & $367^{+18}_{-18}$     & $89^{+4}_{-4}$       & (-0.8,-41.5,-26.9)   \\
      $M05$    &(-4.3,-182.45,169.8)            & 249.3      & 237.9          & 74.4           & (0,-43.9,-25.04)     \\
      $vdM02$  &$(-56^{+36}_{-36},-219^{+23}_{-23},186^{+35}_{-35})$ & $293^{+39}_{-39}$ & $281^{+41}_{-41}$     & $84^{+7}_{-7}$       & (-0.8,-41.5,-26.9)   \\

\hline
\end{tabular}
\caption{Orbital parameters of LMC used in this work. The position
is in $kpc$ and velocities are in $kms^{-1}$ and is measured in
Galactocentric coordinate. References: Gardiner \& Noguchi (1996,
{GN96}), Heller \& Rohlfs (1994, {HR94}), Kallivayalil et al.
(2006a,b, {K1} and {K2}), Mastropietro et al. (2005, {M05}), and van
der Marel et al. (2002, {vdM02}).\label{tmodel} }
\end{center}
\end{table*}

(b){\it Hydrodynamical drag force:} The hydrodynamical friction
force results from the direct collision of the halo gas with the MCs
while passing through the halo. The presence of hot diffused gas in
the Galactic halo in hydrodynamical equilibrium with the dark halo
has been proposed by White and Frenk (1991). To explain some
ionization features discovered in MS, the baryonic halo gas should
have a temperature in order of $\sim 10^6 K$ distributed at a radius
larger than $70 kpc$ \cite{sem03,put03}. Constraints from the
dynamics and the thermal observations propose the density of gas to
be in the range of $10^{-5}$ to $10^{-4} cm^{-3}$. Assuming MCs as a
dense sphere moving through the halo, the drag force on this sphere
is given by:
\begin{equation}
F_{drag}= \frac{1}{2}C_d(R_e)\rho_g v_c^{2}\pi D^2, \label{drag}
\end{equation}
where $C_d(R_e)$ is the coefficient of drag force which is a
function of Reynolds number $R_e$, $\rho_g$, the density of halo,
$v_c$, the relative speed with respect to the gas, and $D$, the size
of the Clouds.
%The Reynolds number is a function of the size and velocity of
%the sphere, as well as the dynamical viscosity, $\eta=
%\frac{1}{3}\rho_g <v_{rms}>\lambda \label{viscosity}$, where
%$<v_{rms}>$ is the mean thermal velocity and $\lambda$ is the mean
%free path of the particles.
%To see the contribution of the drag force on the dynamics of MCs, we
%again compare the time scales of dynamical and hydrodynamical drag
%force.
The time scale of the hydrodynamical friction force can be given by
$T_{drag} \sim M_{MCs}v_c/F_{drag}$. which implies $T_{drag}/T_{dyn}
\sim M_{MCs}/\rho_grD^2$.
%\begin{equation}
%\frac{T_{drag}}{T_{dyn}} \sim \frac{M_{MCs}}{\rho_grD^2},
%\end{equation}
Substituting the parameters as $r\sim 50kpc$, $D \sim 5 kpc$, number
density of halo $\sim 10^{-4.5} atom/cm^3$ and $M_{MCs}\sim10^{10}
M_{\odot}$, we obtain $T_{drag}\gg T_{dyn}$. Figures~(\ref{d1}b) and
~(\ref{d1}c) show the effect of friction forces appeared in equation
(\ref{lmc}) on the orbital motion of MCs. Due to the dynamical
friction, MCs lose a few percent of total energy in each passage.
The effect of the gaseous halo density and size of clouds on the
magnitude of drag force is indicated in Figure~(\ref{d1}). The
apogalactic distance of clouds decreases gradually during the
evolutionary period. Finally, we calculate the dynamics of MCs
including all forces, represented in Figure~(\ref{d1}d). Amongst the
dissipative forces, the dynamical friction force is the dominant
term and can deviate the orbit of MCs from that of frictionless one.

\subsection{Modeling of the Magellanic Stream }
In order to determine the evolution and the formation history of MS,
it is essential to know the spatial location and velocities of MCs
at the present time as the initial condition. As we mentioned before
this initial condition suffers from the uncertainties mainly from
the tangent velocity field of this structure. Different initial
conditions show significantly different orbits and location of the
interaction between LMC and SMC. In this regards, recently an
extended analysis of the parameter space for the interaction of the
Magellanic system with the MW has been done by Ruzicka et al (2007),
using the genetic search algorithm combined with an approximate
restricted N-body simulation (instead of fully self-consistent
simulation). Also in another orbital analyzing, different sets of
initial conditions have been applied to determine the first passage
of LMC around the MW \cite{bes07}. One of the main aims of this work
is seeking the results that are common for different selection of
initial conditions.

We use the six different sets of initial conditions listed in Table
(\ref{tmodel}) and for each set, we find out the best space
parameters of MW halo. Since the obtained parameters depends on the
formation mechanism of the MS, we compare our results with the other
studies which used other formation mechanism.
%It should be mentioned that despite all differences the current
%position of LMC is always close to the Perigalaction where the
%effects of the interaction with the MW become evident.
%In general, the ram pressure force significantly affects on the
%gaseous medium and not on the stars and thus this model seems to be
%more consistent with a gas-only structures and allow for a better
%reproduction of HI column density profile than the tidal model.
For modeling the MS, we follow the continues ram pressure stripping
mechanism introduced by Sofue (1994) including the dynamical and
hydrodynamical friction forces. In that work it is assumed that if
LMC and SMC approach to each other, this sever encounter will most
likely disrupt the two galaxies to form
%The MCs will interact over a
%significant amount of time without catastrophic encounters.
%He suggested that
the Magellanic bridge (a stream of HI gas between the LMC and SMC).
The Magellanic bridge within SMC and LMC then has been stripped off
by the ram-pressure when the MCs was moving through a hot MW halo.
According to the observations, at the present time, the gaseous
stuff is leaving the Magellanic bridge \cite{bru05}. We assume that
the gas is initially distributed inside the MCs. The relative motion
of MCs inside the Galactic halo makes a drag force on the gas of
Magellanic bridge and due to the larger cross section of gas compare
to the stars, the gas exit from the MCs. The released gas slows down
by the friction force and accrete into the galactic disk and makes
the MS. The equation of motion of MS particles can be written as
\begin{equation}
\frac{d^{2}\mathbf{r}_{MS}}{dt^{2}}=\frac{\partial}{\partial
\mathbf{r}_{L}}[\phi_{MCs}(|\mathbf{r}_{MS}-\mathbf{r}_{MCs}|)+
\phi_{G}(|\mathbf{r}_{MS}|)] + \mathbf{f} + F_{drag},
\end{equation}
where $r_{MS}$ and $r_{MCs}$ are the distance of MS and MCs from the
center of Galaxy, $\mathbf{f}$ is the dynamical friction force and
$F_{drag}$ is the hydrodynamical drag force. As a simple model, MS
is considered as a series of spherical clumps with the size of about
$0.5 kpc$ and the mass of $m=3 \times 10^6 M_{\odot}$ \cite{sof94}.
The first part of the MS (i.e. end of tail) is generated at the time
of LMC-SMC close approach about $0.5Gyr$ ago and subsequent clumps
is released from the MCs after this time. Since the dynamical
friction force is proportional to the mass of the object, for these
clumps, we can ignore this term in comparison with the drag force.
Base on such simple model, we were able to produce a large number of
trailing tail across the entire range of halo parameters values. By
comparing the created MS profile with the observed data we find out
the best set of halo parameters. In the next section we calculate
the orbit of MS for two main category of the power-law and
logarithmic models.

\section{Results and discussion }\label{S4}

In this section we calculate the equations of motion of MCs using
the gravitational potential of disk, bulge and dark halo. We extract
the dynamics of MS and obtain the trajectory of MS in various
Galactic models. Then we reproduce the radial velocity distribution
of HI in MS for different initial conditions of MCs (reported in
Table \ref{tmodel}) to compare with the observed data. For each
initial condition, the best fit orbital parameters have been
reported in Table (\ref{ResultL}). We assumed that the LMC and SMC
form a binary system that has been in a slowly decaying orbit about
the MW for roughly a Hubble time. One of the crucial points in
studying the dynamics of MS is the location of the close approach
and the corresponding time which depends on the velocities and
positions of MCs at the present time. For the heavy halos, the
collision happens at the closer distance to the present position of
MCs than the light mass halos.

In what follows we use generic power-law and logarithmic halo models
and extract the best values for the parameters of the model through
the maximum likelihood analysis. We use $\chi^2$ fit to compare the
deviation of the observed radial velocity from the theoretical
predictions.

\subsection{Logarithmic model}

\begin{table*}
\begin{center}
\begin{tabular}{|c|c|c|c|c|c|}
\hline
work                   &$V_C(km/s)$      &         $q$            & $T_{col}[Gyr]$ & $d_{min}[kpc]$& $\chi^2$ \\
 \hline $GN96$         & $180^{+1}_{-1}$ & $0.95^{+0.05}_{-0.02}$ & 0.33           & 5.9           & 7.9      \\
 $HR94$                & $195^{+1}_{-1}$ & $0.55^{+0.07}_{-0.05}$ & 0.97           & 6.7           & 9.8      \\
 $K2$                  & $190^{+3}_{-7}$ & $0.85^{+0.15}_{-0.10}$ & 0.25           & 8.2           & 10.5      \\
 $K1 Mean$             & $195^{+4}_{-5}$ & $1.00^{+0.15}_{-0.07}$ & 0.22           & 7.4           & 8.9     \\
 $M05$                 & $165^{+2}_{-1}$ & $1.05^{+0.07}_{-0.02}$ & 0.46           & 4.0           & 8.6    \\
 $vdM02$               & $200^{+1}_{-4}$ & $0.9^{+0.10}_{-0.02}$  & 0.30           & 10.03         & 8.4      \\

\hline
\end{tabular}
\end{center}
\caption{  The best Parameters of \textsl{logarithmic} model for
various initial conditions. The first column represents the
corresponding reference for the initial condition of MCs. The
parameters in the second and third columns ($V_c$ and $q$ ) have
been introduced in the text. The fourth column, $T_{col}$ is the
last close approach of LMC and SCM, the fifth column $d_{min}$
indicates the distance between LMC and SMC at last close approach.
The last column shows the minimum $\chi^2$ corresponds to each
initial condition. The mean value of $V_c$ and $q$ from the various
initial conditions for MCs is $\bar{q}\sim0.9$ and
$\bar{V_c}\sim187$. \label{ResultL}}
\end{table*}

\begin{figure}[h]
  \begin{minipage}[t]{0.99\linewidth}
  \centering
   \includegraphics[width=90mm,height=60mm]{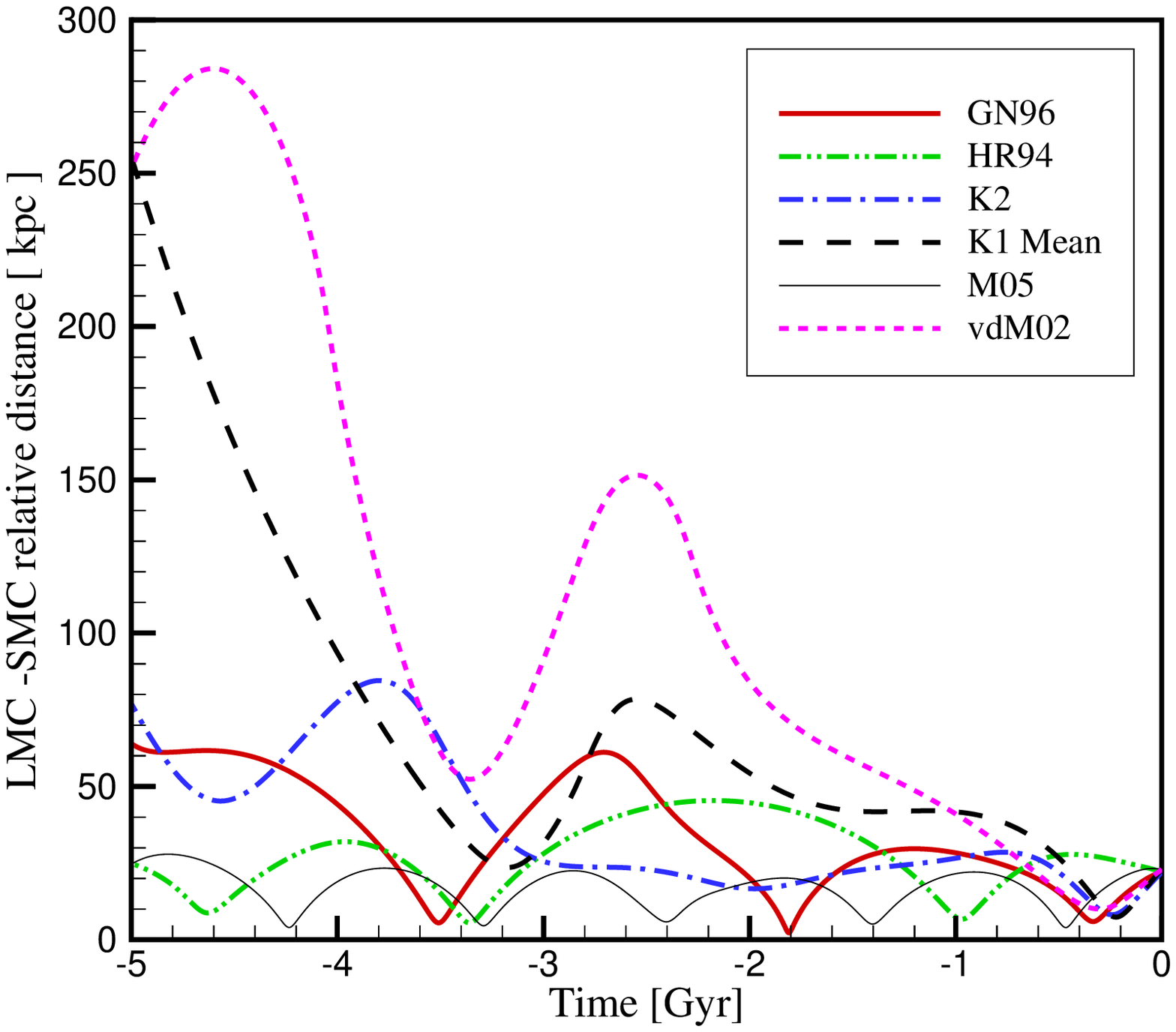}
   \includegraphics[width=90mm,height=60mm]{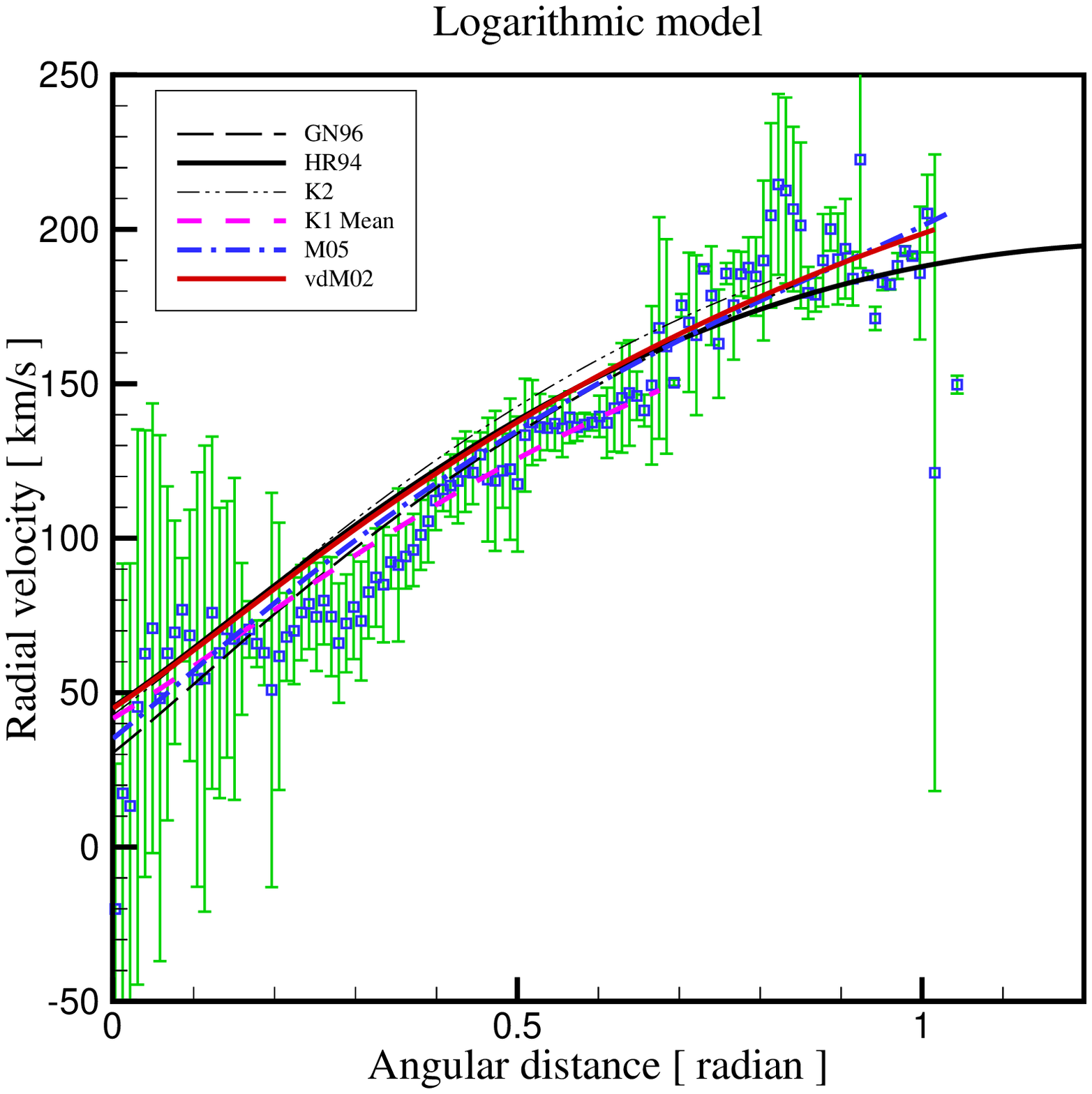}
   \caption{{Relative distance of LMS and SMC
as a function of time (top) and radial velocity versus angular
distance along the $MS$ (bottom) for the best-fit
\textsl{logarithmic} halo potential. Open squares show the observed
radial velocity and curves correspond to that of best fit with
various initial conditions. }\label{vrL} }
  \end{minipage}%
%  \begin{minipage}[t]{0.495\textwidth}
%  \centering
%   \includegraphics[width=80mm,height=80mm]{fig5.eps}
%  \caption{{\small Relative
%distance of LMS and SMC as a function of time for the various
%initial conditions. The MW dark halo is modeled as a
%\textsl{logarithmic} potential and the corresponding best-fit values
%of parameters have been considered for various initial conditions.}}
%  \end{minipage}%
  %\label{DisL}
\end{figure}

\begin{figure}{}
\begin{center} \resizebox{13cm}{!}{\includegraphics[width=190mm,height=190mm]{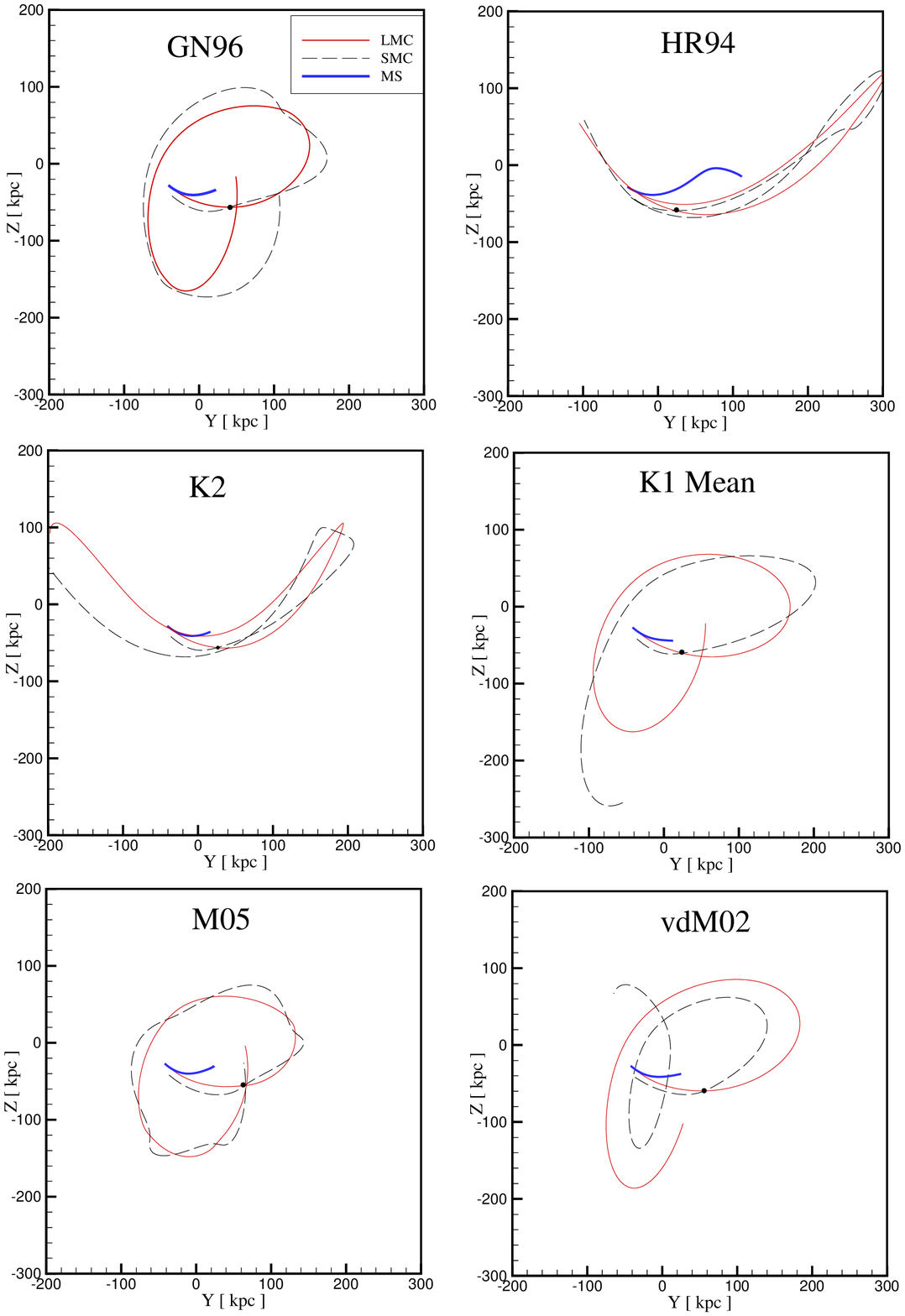}}
\caption{$Y-Z$ plane orbital path of LMC (red solid line), SMC
(dashed line) and MS (blue solid line) for the best-fit
\textsl{logarithmic} halo potential with various initial conditions.
The orbital path of the LMC is not significantly affected by the
SMC, however SMC's path is deviated by LMC. MW is located at
$(0,0)$. The black circle denotes the position of last close
approach of LMC and SMC. The duration of orbital evolution is $5
Gyr$.}\label{orbitL}
\end{center}
\end{figure}

We use radial velocity of MS to compare the theoretical model with
the observed data. The radial velocity of MS is calculated by
averaging over the data at each point of MS along the line of sight.
The corresponding error bar results from the velocity dispersion of
the structure which results from the stochastic motions of gas.
% We use the
%generic power-law and logarithmic halo models and extract the best
%values for the parameters of the model, through the maximum
%likelihood analysis. We use $\chi^2$ fit to compare the deviation of
%the observed radial velocity from the theoretical predictions.
Figures~(\ref{vrL}) represents the relative distance of LMC and SMC
for six different initial conditions and the radial velocity of MS
as a function of angular location for the best free parameters of
the model. It should be noted that radial velocity is measured with
respect to an observer located at the center of Galaxy. The zero
angular distance corresponds to the position of $MS I$ (the nearest
part of MS to MCs).

For each set of halo parameters, the radial velocity profile of HI
clumps in MS is numerically calculated and compared with the
observed radial velocity profile. The analysis repeated for various
initial conditions of the clouds. The best value of model parameters
($q$ and $V_c$) for various initial conditions are given in
Table~\ref{ResultL}. Except {HR94} which prefers more oblate halo,
for the other sets of initial conditions, better agreement between
the model and observational data is achieved for nearly oblate or
spherical halo. The mean value of the best fit parameters, averaging
over the different sets of initial conditions are $\bar{q}\sim0.9$
with $\bar{V_c}\sim187kms^{-1}$. These values are in agreement with
the flattening of the MW halo potential obtained by Helmi (2004) and
are also compatible with the implied value of circular velocity of
MW \cite{mcg08,xue08}.

%In Figure~(\ref{orbitL}) also we plot the orbital path of the LMC
%and SMC with MS for six different set of initial conditions. Finally
%Table ~(\ref{ResultL}) represents the best free parameters for the
%logarithmic halo model.

%Figure~\ref{vrL} compares the best fit theoretical curves of radial
%velocity profile for each initial condition with the observed data
%(Br\"{u}ns et al. 2005) and also shows the time evolution of
%relative distance of LMC and SMC for the last 5 Gyr.

Finally we present the projected trajectory of LMC, SMC and MS in
Figure~(\ref{orbitL}) for various initial conditions. The MCs move
on rosset-shape orbits except the case of {HR94} and {K2} which have
the banana-shape trajectory. The apogalactic distance of LMC
decreases gradually during the evolutionary period, which reflects
the effect of dynamical friction. According to
Figure~(\ref{orbitL}), for two initial conditions represented by
{K2} and {K1-mean}, the spatial extension of MS is shorter than the
observed size, while for the case of {HR94} it is longer than the
observed size. This result arises from the longer time for the last
close approach ($T_{col}$) in {HR94}. The minimum relative distance
of the LMC and SMC at the last close approach ($d_{min}$) and the
corresponding time is calculated and presented in
Table~\ref{ResultL}.

\subsection{Power-law model}

Following the logarithmic model we calculate the radial velocity
distribution of MS in the power-law galactic model and compare it
with the observed data. In the power-law model we have $4$ free
parameters: the asymptotic velocity, $V_a$, the halo flattening,
$q$, the core radius, $R_c$ and the rising or falling parameter of
rotation curve $\beta$. The best value of parameters for each
initial conditions are given in Table~\ref{ResultP}. The mean value
of the best fit parameters averaging over the different set of
initial conditions are $\bar{q}\sim0.85$, $\bar{V_a}\sim168$,
$\bar{R_c}\sim17$, and $\bar{\beta}\sim-1.7$. Again similar to the
logarithmic halo model, a better agreement with observational data
is generally achieved for nearly oblate or spherical ($q\leq1$) halo
model.
%This result is in agreement with that of logarithmic model
%with $\bar{q}\sim0.9$ \cite{hel04,ruz07}.

\begin{table*}
\begin{center}
\begin{tabular}{|c|c|c|c|c|c|c|c|}
\hline
work           & $V_a[km/s]$     & $R_c [kpc]$   & $\beta$                & $q$                   & $T[Gyr]$& $d[kpc]$& $\chi^2$ \\
\hline$GN96$   & $185^{+2}_{-2}$ & $17^{+2}_{-4}$& $0.10^{+0.06}_{-0.20}$ & $0.95^{+0.05}_{-0.03}$& 0.33          & 5.9           & 7.86     \\
      $HR94$   & $165^{+4}_{-5}$ & $13^{+6}_{-4}$& $-0.20^{+0.14}_{-0.04}$& $0.60^{+0.12}_{-0.04}$& 0.76          & 4.8           & 9.5      \\
      $K2$     & $160^{+5}_{-6}$ & $13^{+4}_{-2}$& $-0.20^{+0.02}_{-0.09}$& $0.80^{+0.18}_{-0.08}$& 0.23          & 7.5           & 8.7      \\
      $K1 Mean$& $165^{+7}_{-10}$& $17^{+6}_{-7}$& $-0.20^{+0.10}_{-0.10}$& $0.75^{+0.17}_{-0.10}$& 0.22          & 6.5           & 8.5      \\
      $M05$    & $145^{+10}_{-2}$& $21^{+3}_{-4}$& $-0.30^{+0.06}_{-0.10}$& $1.00^{+0.04}_{-0.03}$& 0.41          & 2.1           & 7.2      \\
      $vdM02$  & $190^{+2}_{-2}$ & $21^{+3}_{-3}$& $-0.20^{+0.04}_{-0.04}$& $1.00^{+0.03}_{-0.06}$& 0.28          & 9.45          & 7.5      \\

\hline
\end{tabular}
\end{center}
\caption{The best parameters of \textsl{power-law} model for each
sets of initial condition. The parameters of this model is
introduced in equation (\ref{phiP}). The mean value of the best fit
parameters averaging over the different set of initial conditions
are $\bar{q}\sim0.85$ and $\bar{V_a}\sim168$, $\bar{R_c}\sim17$,
$\bar{\beta}\sim-1.7$.
\label{ResultP} }
\end{table*}

\begin{figure}[h]
  \begin{minipage}[t]{0.99\linewidth}
  \centering
  \includegraphics[width=90mm,height=60mm]{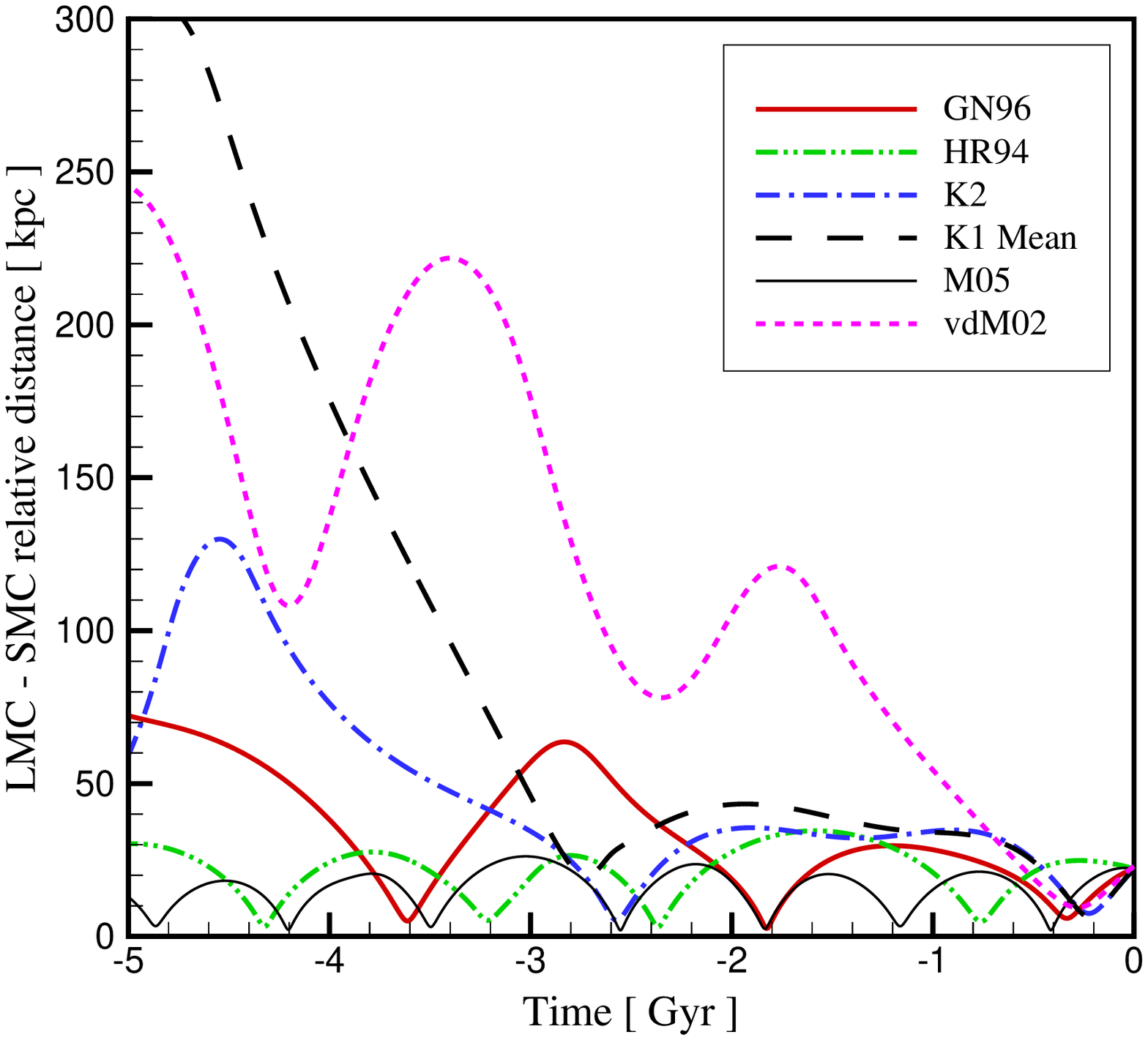}
  \includegraphics[width=90mm,height=60mm]{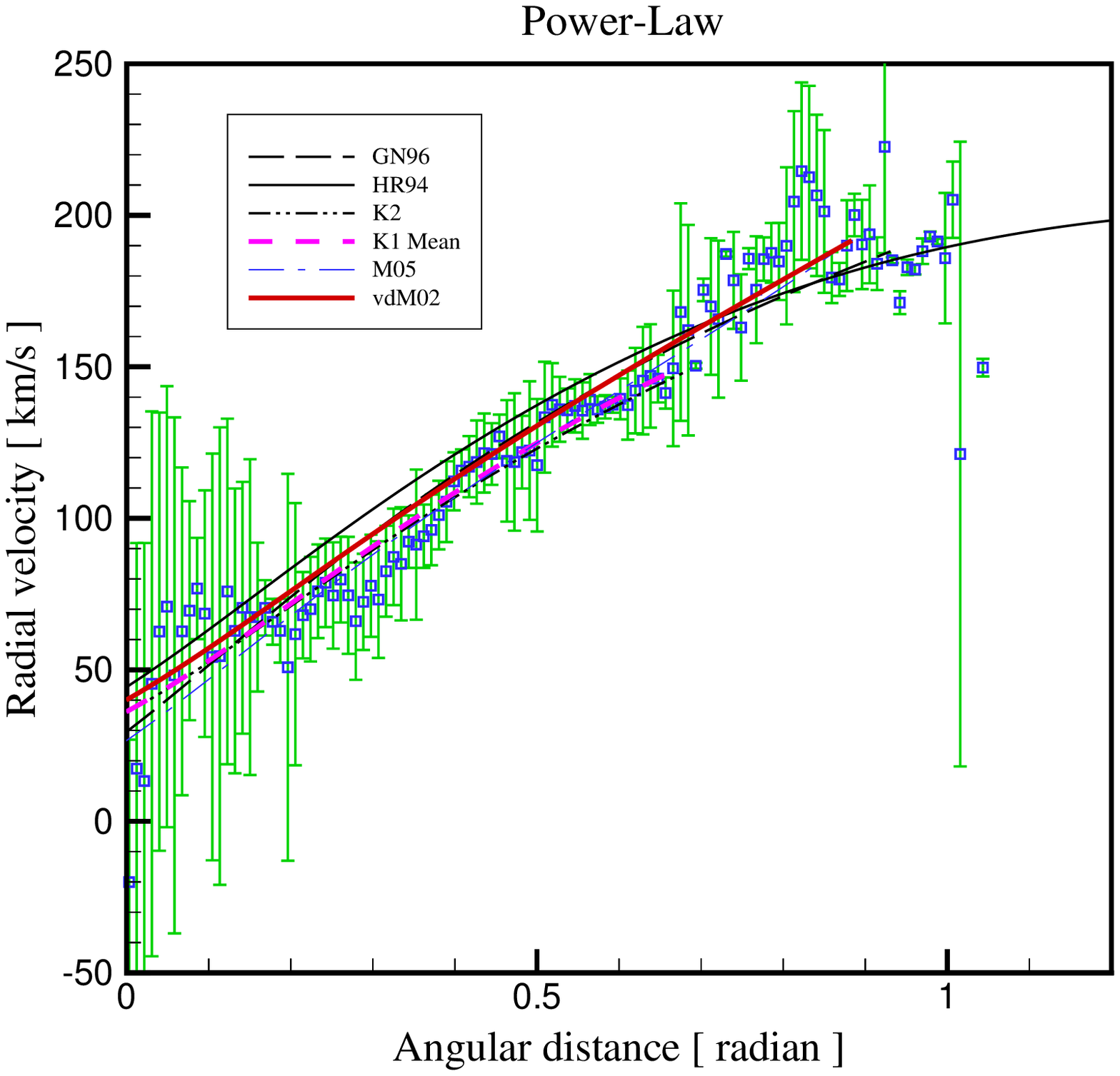}
\caption{Relative distance of LMS and SMC as a function of time
(top) and radial velocity versus angular distance along the $MS$
(bottom) for the best-fit \textsl{Power-law} halo potential. Open
squares show the observed radial velocity and curves correspond to
that from the best fit with various initial conditions.}\label{vrP}
  \end{minipage}
\end{figure}

\begin{figure}{}
\begin{center} \resizebox{13cm}{!}{\includegraphics[width=200mm,height=200mm]{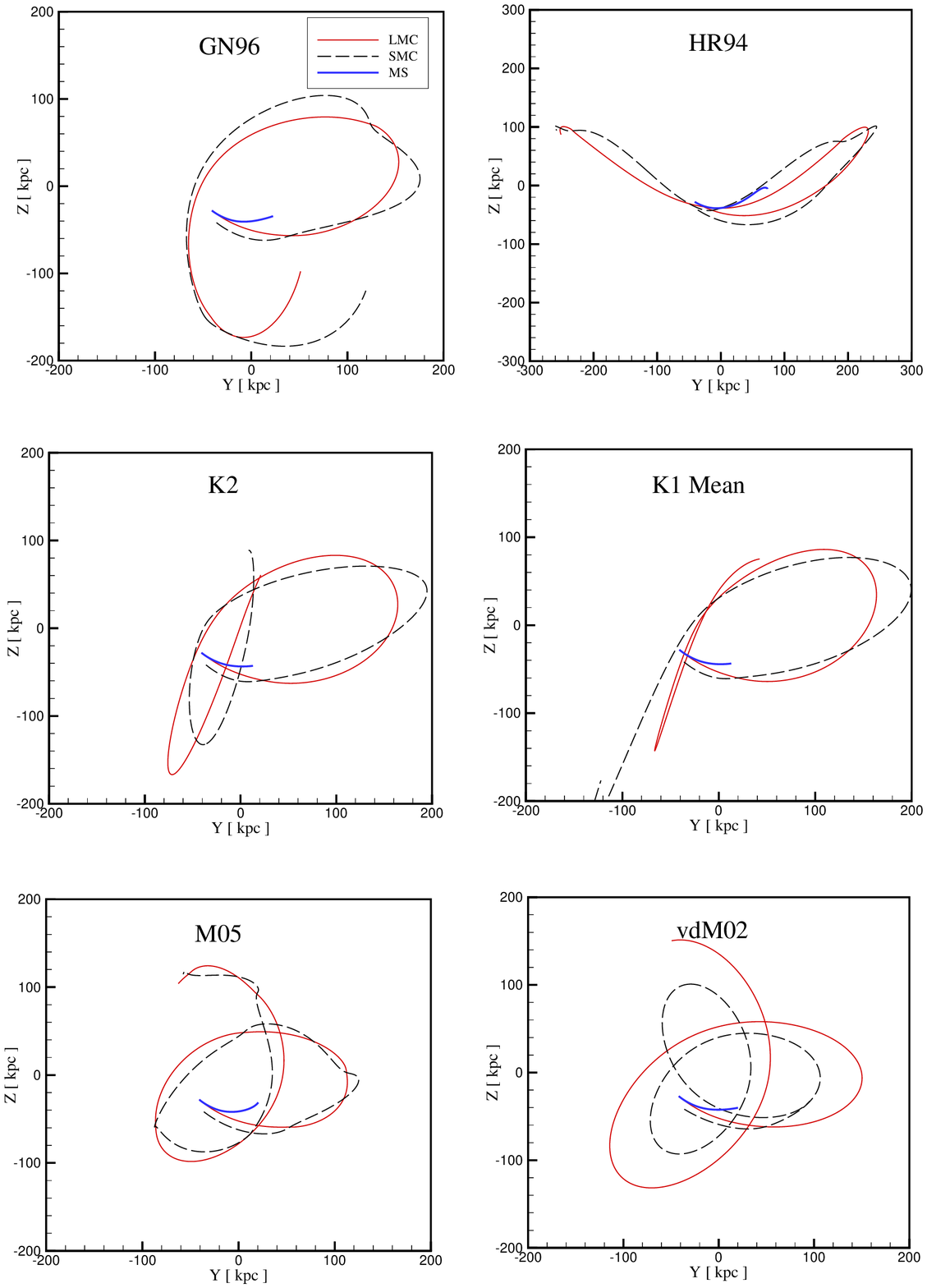}} \caption{
The $Y-Z$ plane orbital path of LMC (red solid line), SMC (dashed
line) and MS (blue solid line) for the best-fit \textsl{power-law}
halo potential with various initial conditions. The orbital path of
the LMC is not significantly affected by the SMC, however SMC's path
is deviated by LMC. MW is located at $(0,0)$. The black circle
denotes the position of last close approach of LMC and SMC. The
duration of orbital evolution is $5 Gyr$.
%The $Y-Z$ plane orbital
%path of LMC (red solid line), SMC (dashed line) and MS (blue solid
%line) for the best-fit \textsl{power-law} halo potential in various
%initial conditions. The MW center is located at $(0,0)$. The black
%circle denoted the position of last close approach of LMC and SMC.
%The time of orbital evolution is $5 Gyr$. The best-fit halo
%potential parameters are presented in table \ref{ResultP}.
 }\label{orbitP}
\end{center}
\end{figure}

Figure~(\ref{vrP}) shows the time variation of the relative distance
of LMC and SMC for the last 5 Gyr and radial velocity in Galactic
frame as a function of angular separation of MS with respect to MSI.
Different curves corresponds to various initial conditions of the
MCs. The projected trajectory of LMC, SMC and MS for the best values
of the parameters of model is shown in Figure~(\ref{orbitP}).
Similar to the logarithmic model, in the case of {HR94}, the
generated MS extends longer than the observation. From the equations
of motion, the minimum relative distance of the LMC and SMC at the
last close approach and the corresponding time is also calculated
(see Table~\ref{ResultP}).

%  \begin{minipage}[t]{0.495\textwidth}
%  \centering
%   \includegraphics[width=80mm,height=80mm]{fig15.eps}
%  \caption{{\small Relative
%distance of LMS and SMC are plotted as a function of time. The MW
%dark halo is modeled as a \textsl{power-law} potential. For each set
%of initial conditions, the corresponding best-fit values of model
%parameters $(V_a, q, R_C, \beta)$ have considered.}
%  \end{minipage}%

%\begin{figure}{}
%\begin{center}
%\resizebox{8cm}{!}{\includegraphics{fig14.eps}} \caption{ The radial
%velocity $v_r$ vs the angular distance $\theta$ along the $MS$ with
%respect to MSI for the best-fit \textsl{Power-law} halo potential.
%The data observed by Br\"{u}ns et al. 2005. Different lines
%corresponds to the various intial conditions for the MS.}\label{vrP}
%\end{center}
%\end{figure}
%
%\begin{figure}{}
%\begin{center}
%\resizebox{8cm}{!}{\includegraphics{fig15.eps}} \caption{Relative
%distance of LMS and SMC plotted as a function of time for the
%various initial condition. The MW dark matter halo is modeled as a
%\textsl{Power-law} potential and the corresponding best-fit values
%of model parameters, $(V_a, q, R_C, \beta)$, have considered for
%each set of initial conditions. }\label{DisP}
%\end{center}
%\end{figure}

\begin{figure}{}
\begin{center}
\resizebox{12cm}{!}{\includegraphics{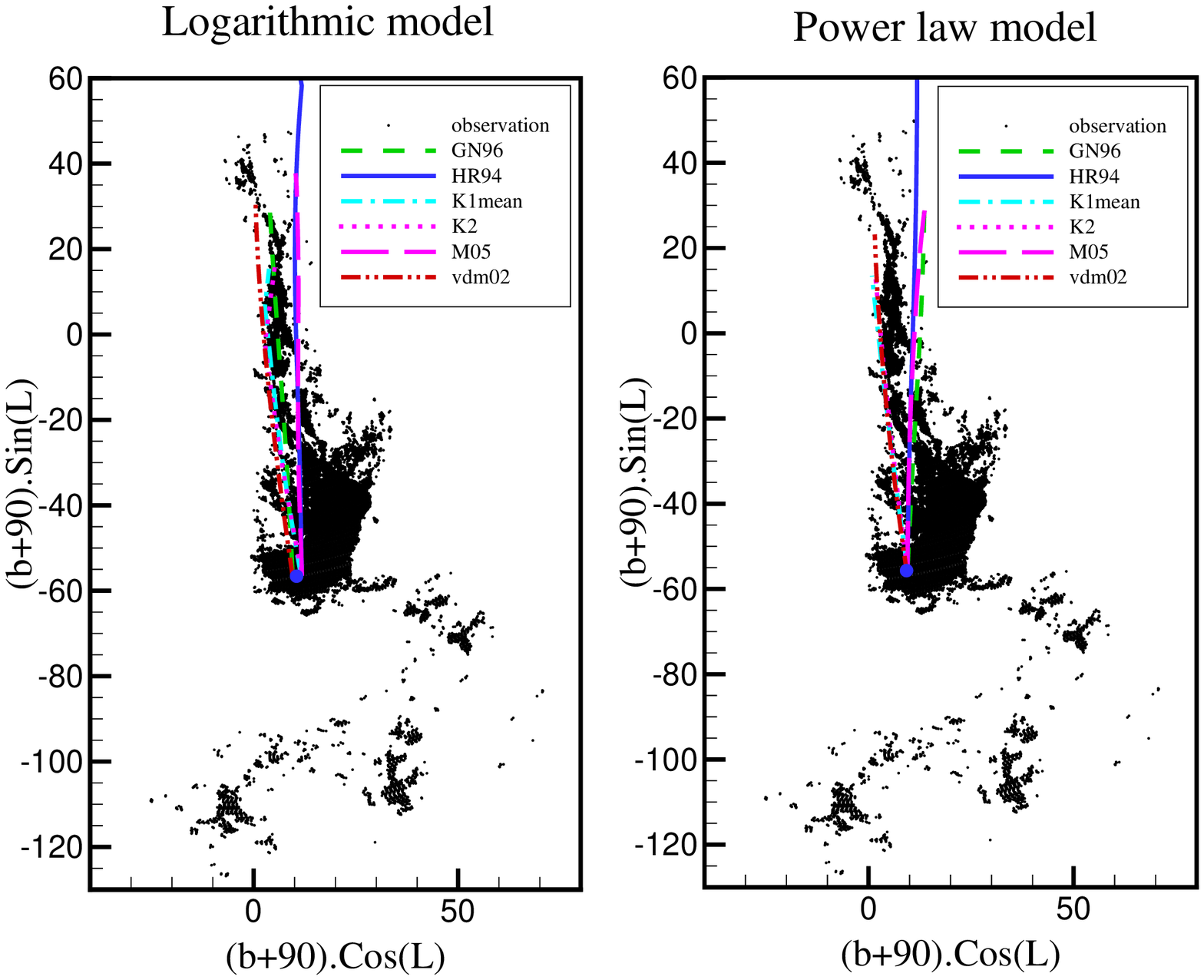}} \caption{ HI
distribution of MS from the data \cite{bru05} is plotted as a polar
projection for both logarithmic and power law model. The axis are
the galactic longitude and latitude as defined in the axis label.
The blue dot indicates the location of the LMC. The colored lines
are the projection of spatial distribution of the MS which
theoretically determined by several initial conditions under
logarithmic and power-law MW model. Note that the different initial
conditions almost does match the data. The HR94 result is
indistinguishable from that of the M05 result and both deviate from
the location of the tip of MS. The similar situation occur for the
vdm02, K1mean and K2 velocities. \label{lb}}
\end{center}
\end{figure}

In order to visualize the compatibility of simulated MS with the
observational distribution of the stream, we plot the projected
spatial distribution of MS corresponding to the different initial
conditions in Figure~(\ref{lb}). We find little deviation in the
projected distribution of MS with the theoretical predictions.
%However this compatibility is better for in the case of
However, there is a significant difference between the observation
and the theoretical prediction in the case of HR94 and M05 for both
halo models. Although the GN96 initial condition could fit with the
MS in logarithmic model, it no longer traces the MS in power-law
model. Figure~(\ref{lb}) shows that in the logarithmic model the
projected MS using the GN96, vdm02, K1mean and K2 initial conditions
are nearly indistinguishable and almost overlaying HI data of MS.
The similar results valid for power-law model except for the case of
GN96.

\section{Conclusion}\label{S5}
%=====================================================================
Summarizing this work, we studied the formation of Magellanic Stream
as a consequence of interaction of the Magellanic Clouds and MW. We
used a continuous ram-pressure model for simulating MS. The dynamics
of MS is used to put constrain on the shape of Galactic halo. In
this model the particles of halo sweep the hydrogen gas of the
Magellanic Clouds while moving through the halo.
%We presented a simple model for the formation and evolution of
%the MS.
At the first step we reconstruct the orbital motion of the
Magellanic Clouds as a three body system of Galaxy, LMC and SMC
taking into account the initial condition of last two objects at the
present time. In this scenario, at the last close approach of LMC
and SMC a part of their gas has been ejected from the system to form
Magellanic stream. While the Magellanic Clouds gradually spiral
towards the Galaxy, the stream of gas releases from this system and
makes Magellanic Stream. We showed that the drag force of halo on
the stream is larger compare to that on the Magellanic Clouds while
the dynamical friction force on Magellanic Clouds is larger than on
Magellanic Stream. These friction forces causes the orbit of these
structures to spiral faster toward the Galaxy. Using this simple
model for generating MS, we were able to find the best parameters of
halo model, comparing the radial velocity of Stream with the
theoretical model.

In order to see the effect of different shape of Galactic
potentials, we took two generic logarithmic and the power-law
potentials for the Galactic halo and calculate the dynamics of
Magellanic Stream for these potentials. Since there is a large
uncertainty in the initial conditions of MCs at the present time, we
applied six different initial conditions reported in the literature
\cite{bes07}. Finally we compared the numerical results of radial
velocity profile of MS with the observed data. Using the likelihood
analysis we found that Galactic halo is nearly oblate almost in all
the initial conditions for the two galactic halo models.
Furthermore, the preferred value for the flattening parameter in our
analysis is in agreement with the recent analysis which has employed
a tidal model for the MS and logarithmic potential for the Galactic
halo \cite{ruz07}. In addition, the corresponding value for circular
velocity is compatible with the recent measurement based on SDSS
data \cite{xue08} and is in agreement with the rotation curve
analysis \cite{mcg08}.

%It should be noted that this result is independent of selected
%initial condition and is valid for both logarithmic and power-law
%halo model.

%\begin{acknowledgements}
We would like to thank C. Br\"{u}ns provided us recent data of
Parkes HI survey of Magellanic System and his useful comments.
%\end{acknowledgements}

\end{document}